\documentclass[12pt]{spieman}  
\usepackage[]{graphicx}
\usepackage{setspace}
\usepackage{tocloft}

\title{IO:I: A Near-Infrared Camera for the Liverpool Telescope} 

\author{R. M. Barnsley\supscr{a}, H. E. Jermak\supscr{a}, I. A. Steele\supscr{a}, R. J. Smith\supscr{a}, S. D. Bates\supscr{a}, C. J. Mottram\supscr{a}}

\affiliation{\supscrsm{a}Astrophysics Research Institute, Liverpool John Moores University, 146 Brownlow Hill, Liverpool, UK, L3 5RF}

\cftpagenumbersoff{figure}
\cftpagenumbersoff{table} 
\begin{document} 
\maketitle 

\begin{abstract}
IO:I is a new instrument that has recently been commissioned for the Liverpool Telescope, extending current imaging capabilities beyond the optical and into the near infrared. Cost has been minimised by use of a previously decommissioned instrument's cryostat as the base for a prototype and retrofitting it with Teledyne's 1.7$\mu$m cutoff Hawaii-2RG HgCdTe detector, SIDECAR ASIC controller and JADE2 interface card. In this paper, the mechanical, electronic and cryogenic aspects of the cryostat retrofitting process will be reviewed together with a description of the software/hardware setup. This is followed by a discussion of the results derived from characterisation tests, including measurements of read noise, conversion gain, full well depth and linearity. The paper closes with a brief overview of the autonomous data reduction process and the presentation of results from photometric testing conducted on on-sky, pipeline processed data.
\end{abstract}


{\noindent \footnotesize{\bf Address all correspondence to}: Rob Barnsley, Astrophysics Research Institute, Liverpool John Moores University, 146 Brownlow Hill, Liverpool, UK, L3 5RF; Tel: +44 151 231 2934; E-mail:  \linkable{R.M.Barnsley@ljmu.ac.uk}}

\begin{spacing}{1}   

\section{Introduction}
\label{sec:intro} 
The Liverpool Telescope\cite{2004SPIE.5489..679S}\footnote{http://telescope.livjm.ac.uk/} (LT) is a fully robotic 2m telescope sited on the Canary Island of La Palma. It is an unmanned facility insomuch that it does not require an observer during operations. Instead, observations are queued by an autonomous scheduler. The scheduler provides the telescope with a ranked list of suitable observing programmes based on current and predicted future meteorological data, programme priority and other queue optimisation conditions. A scheduled observation can be overriden at any time by a Target of Opportunity (ToO), making the LT particularly well suited to time-domain observational astronomy where its ability for rapid response has been a key asset.

The LT is a multi-purpose facility hosting up to nine instrument simultaneously. The current instrument suite includes a 10$^{\prime}$x10$^{\prime}$ optical imager (IO:O), an integral-field spectrograph\cite{2004AN....325..215M, 2012AN....333..101B} (FRODOSpec), a fast-readout camera\cite{doi:10.1117/12.787889} (RISE), a polarimeter\cite{2012SPIE.8446E..2JA} (RINGO3) and a low resolution, high throughput spectrograph\cite{doi:10.1117/12.2055117} (SPRAT), all of which are available for use on any single night. IO:I extends the LT's imaging capabilities up to a near infrared (NIR) wavelength of $\sim$1.7$\mu$m where the instrument will enable science to be delivered in, although not limited to, the following key areas:

\begin{itemize}\itemsep-1pt
  \item \textbf{Supernovae type Ia (SNe Ia)}. Measurement of NIR light curves for nearby SNe Ia will help to improve our understanding of the physics of SNe Ia explosions, as well as making a contribution to increasing the accuracy of SNe Ia as cosmological distance measures.
  \item \textbf{Gamma-Ray Bursts (GRBs)}. Early time followup of the highest redshift GRBs ($z>6$), including measurements of line of sight dust extinction, will allow for better determination of the physical parameters of these events.
  \item \textbf{Active Galactic Nuclei (AGN)}. Investigations into the variability of the various classes of AGN will allow us to probe the central engines of these objects, helping to discriminate between competing models.
  \item Other areas of both time-domain (e.g. pixel microlensing, variability surveys) and non time-domain (e.g. metallicity gradients in galaxies and followup up X-ray cluster surveys) astronomy, proposals for which have already been solicited and undertaken.
\end{itemize}

\section{Instrument Overview}\label{sec:instrument_setup}
\begin{figure}[ht]
  \begin{center}
    \begin{tabular}{c}
      \includegraphics[height=16cm]{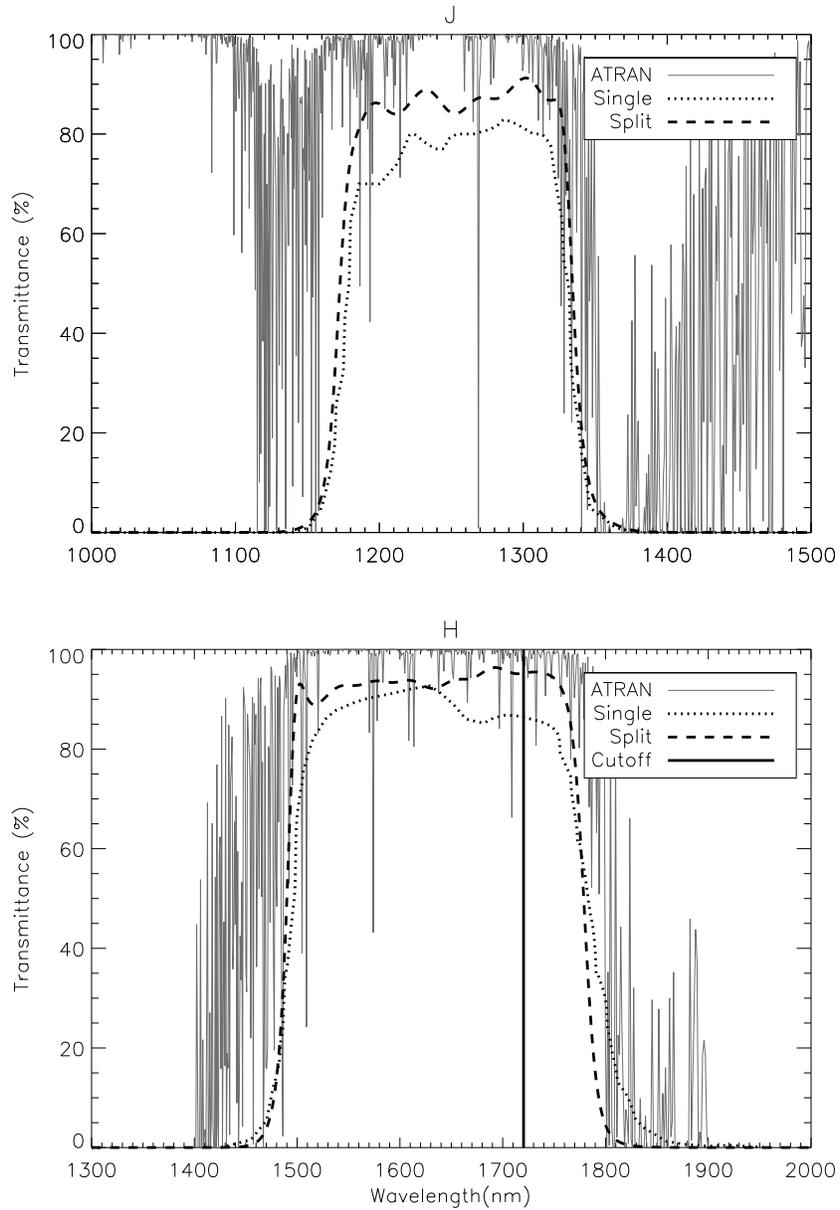}
    \end{tabular}
  \end{center}
  \caption 
  { \label{fig:filters}
  Single and split filter transmittance with modelled atmospheric window derived from ATRAN\cite{1992nstc.rept.....L}. Top: J band. Bottom: H band. The detector cutoff, defined as the point at which the QE drops below 50\%, is shown for the H band.} 
\end{figure}

IO:I is positioned at the Cassegrain focus of the telescope where, without additional powered optics, the plate scale delivered at the focal plane is 97$\mu$m/$^{\prime\prime}$ (2m diameter mirror, f/10 beam focal ratio). With a pixel pitch of 18$\mu$m, the total FOV of the 2048x2048 pixel detector is 6.27$^{\prime}$x6.27$^{\prime}$. The quantum efficiency of the detector is over 50\% between 0.8 and 1.72 $\mu$m (peak 86\% at 1.5$\mu$m), although the usable range within this window is limited by atmospheric transmission (see Figure \ref{fig:filters}). 

The detector itself is a HgCdTe Hawaii-2RG (H2RG) and is controlled via the SIDECAR ASIC controller and JADE2 interface card, all provided by Teledyne. Connected to the cryostat are two closed-cycle cooling units (IGC Polycold Cryotigers), each charged with PT13 gas, providing a total cooling capacity of 10W at a minimum temperature of 80K. Inside the cryostat are two corresponding cold heads. One of these cold heads is connected to the detector/SIDECAR mounting block. The other is attached to the floor of the radiation shield housing, shifting the bulk of the 300K blackbody radiative load incident from the warm cryostat chassis onto this thermal path. By thermally isolating the two paths, the majority of the cooling capacity provided by a single Cryotiger is available to cool the detector and SIDECAR only.

Both J and H single filters have been procured, although the instrument has no filter wheel to allow robotic selection between the two. Following the requirements outlined by science proposals already received for this instrument, the H band filter has been preferred over the J. Changing between these various configurations is not a user selectable option and would require the instrument to be taken off telescope. As a possible future alternative, 3$^{\prime}$x6$^{\prime}$ J and H filters have also been procured, giving the option to split the field-of-view (FOV) of the detector across the filters. With an appropriate dithering strategy, the split configuration would allow the user to observe with both filters in a single observation sequence at the cost of reduced field size. For example, concurrent J and H observations would be obtainable by alternately offsetting the telescope so that when the target is being observed in one band, sky is simultaneously being observed in the other. 
 
The wavelength coverage for each of the filters is shown in Figure \ref{fig:filters}. Flux in the H band will be partially attenuated by the intrinsic 1.72$\mu$m cutoff wavelength of the detector. This cutoff was chosen to restrict the additional complexity required when observing further into the infrared, especially with warm optics and non-IR optimised baffling.

\subsection{Mechanical}

\begin{figure}[ht]
  \begin{center}
    \begin{tabular}{c}
      \includegraphics[height=6.0cm]{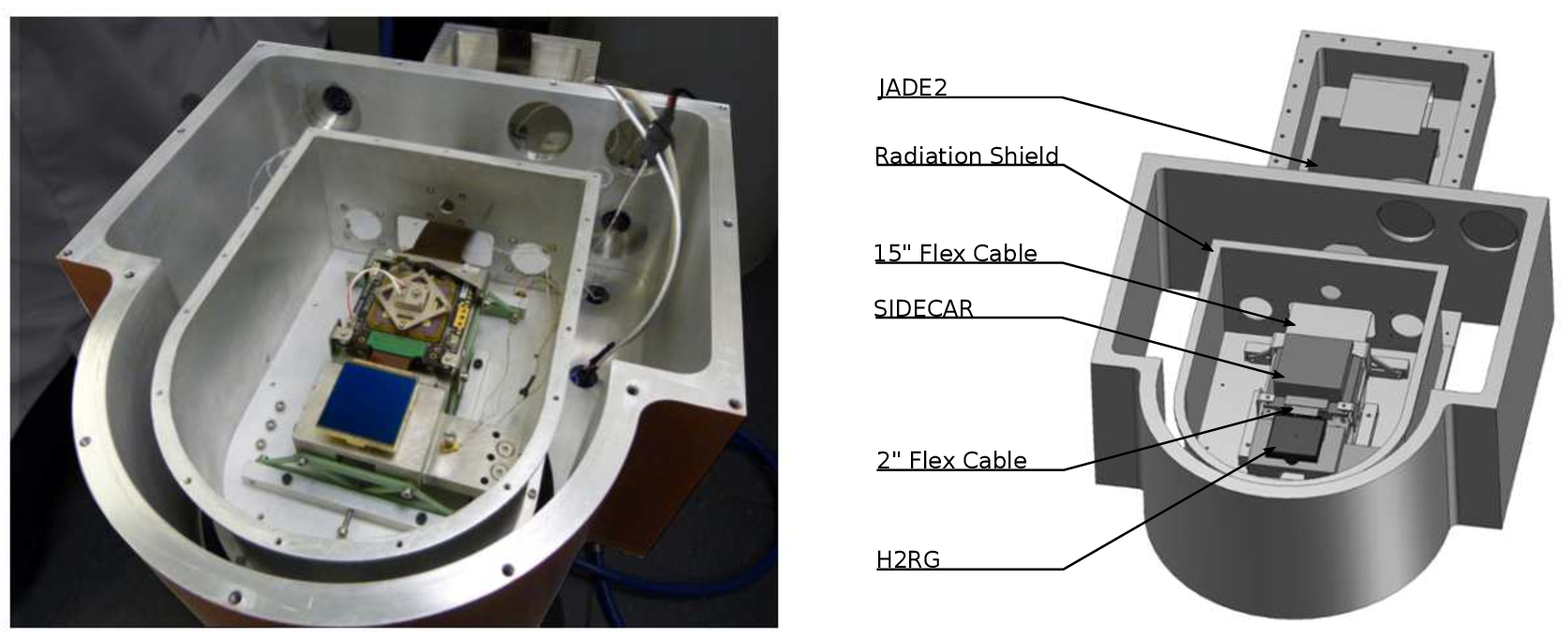}
    \end{tabular}
  \end{center}
  \caption 
  { \label{fig:IOI_1}
  Left: Dewar with bottom and top cover removed showing the detector and SIDECAR. The enclosure housing the JADE2 is located towards 
  the top of the figure. Right: Mechanical drawing of the cryostat with lid/floor and filter box lid removed. The JADE2 is mounted upside-down so that the extent of the resulting extrusion from the main 
  cryostat chassis is reduced.} 
\end{figure}

\begin{figure}[ht]
  \begin{center}
    \begin{tabular}{c}
      \includegraphics[height=6.5cm]{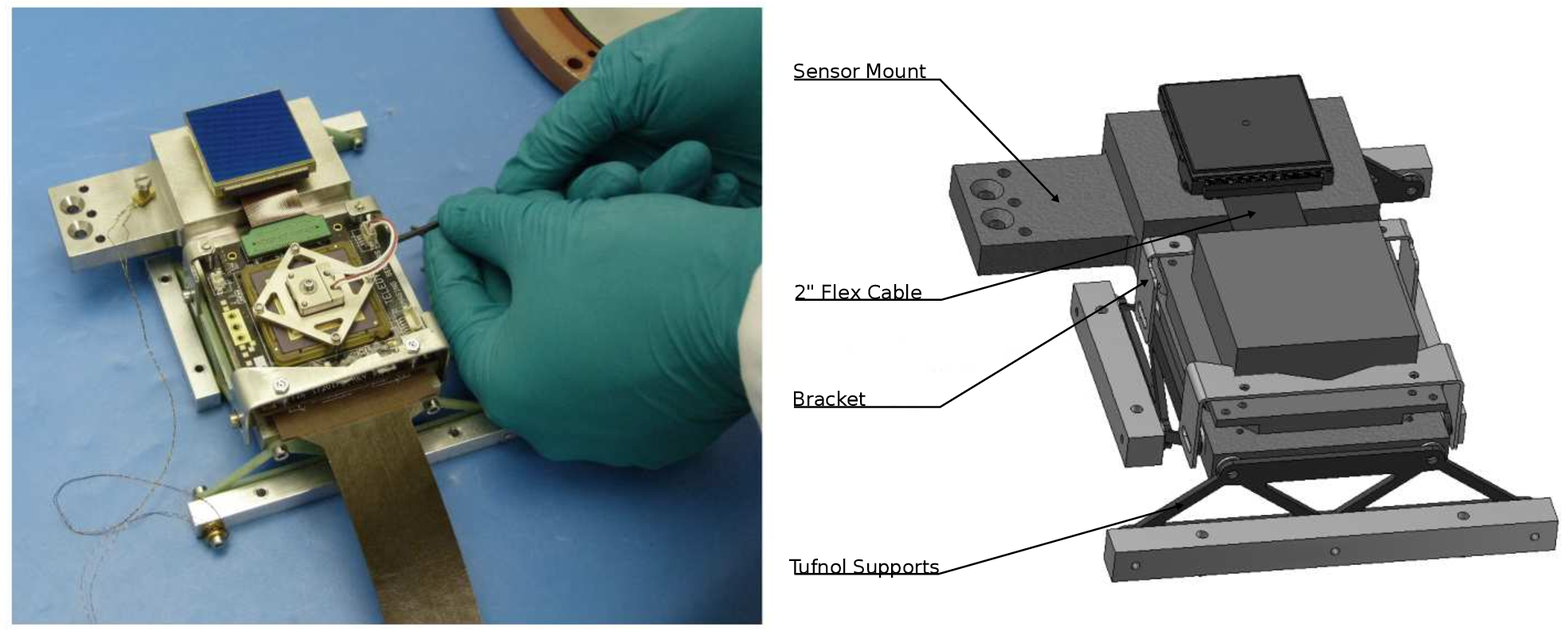}
    \end{tabular}
  \end{center}
  \caption 
  { \label{fig:invar_block}
  Left: Detector/SIDECAR and mounting block. The temperature sensor is visible towards the top left, and is wound around a spool acting as a thermal anchor. Right: Schematic drawing of the mounting block with features labelled. The adjustable mounting bracket allows the SIDECAR to be moved towards/away from the detector, allowing some degree of flexibility when fitting the 2$^{\prime\prime}$ flex cable between the two. The section of the block upon which the temperature sensor and resistor are mounted is connected via a copper block to one of the Cryotiger cold heads.
  }
\end{figure}

Figure \ref{fig:IOI_1} shows the prototype cryostat, recycled from the LT's previous infrared instrument (SupIRCam) with only minor modifications. The H2RG detector and SIDECAR are housed inside an aluminium box which acts as a radiation shield. They are both are mounted upon an invar block, used for its low coefficient of thermal expansion, which in turn is screwed to the base of the box by four thermally insulating Tufnol supports (see Figure \ref{fig:invar_block}). This setup decouples the thermal paths from the two Cryotiger cold heads. The lid of the box holds a 58.5x58.5x5.5mm filter/blank central to the optical axis.

This design closely follows that of RATIR\cite{2012SPIE.8453E..1OF}, in that instead of potting the 15$^{\prime\prime}$ flex cable connecting the (cold) SIDECAR and the (warm) JADE2 through the cryostat wall, the JADE2 is instead kept under vacuum inside the cryostat. In this configuration, both the power and data lines are connected to to a hermetically sealed MIL-Spec socket, with the data transferred externally over a multimode OM1 fibre optic cable to the instrument control computers (ICCs) located off the telescope.

\subsubsection{Additional Baffling}\label{sss:additional_baffling}

\begin{figure}[ht]
  \begin{center}
    \begin{tabular}{c}
      \includegraphics[height=9cm]{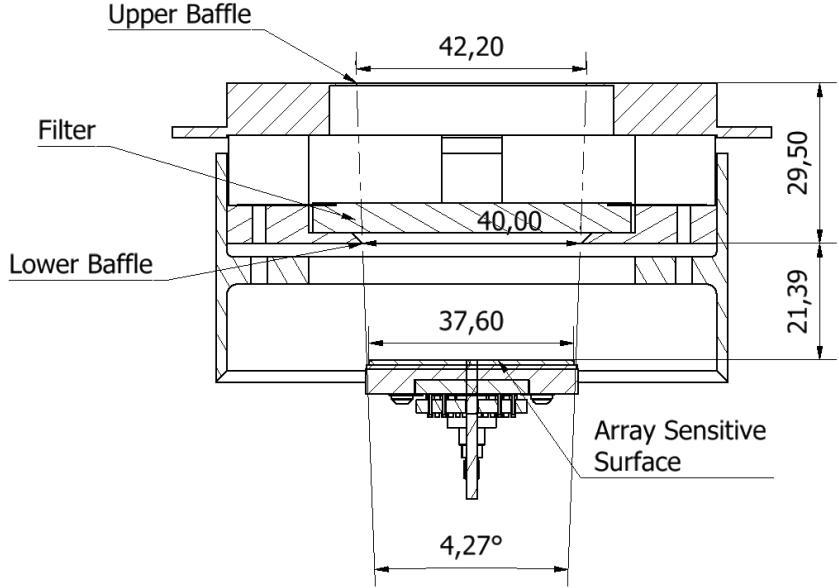}
    \end{tabular}
  \end{center}
  \caption 
  { \label{fig:baffle}
  Section view of the baffle. Light enters from the top. The detector only ``sees'' the warm science fold mirror, located at a distance of 570mm from the detector surface. Units, where unspecified, are in mm.
  }
\end{figure}

Immediately obvious in initial testing was a gross light leak, visible as a $>$1 kADU/s gradient across the array in up-the-ramp (UTR) sequences taken with a cooled blank installed. In UTR mode, the array is read out non-destructively at successive integration times throughout an exposure. Upon further examination, this gradient was thought to arise from i) light incident obliquely on the cryostat window, and/or ii) thermal/LED emission from the JADE2 card. To resolve the issue, a baffle tube was added to the underside of the radiation shield lid, as well as adding two plates to lightly clamp the 15$^{\prime\prime}$ flex cable connecting the SIDECAR to the JADE2 as it feeds through the radiation shield. The baffle tube and corresponding light path to the detector are shown diagrammatically in Figure \ref{fig:baffle}. The telescope uses only optical baffles, and has not been optimised for the NIR.

\begin{figure}[ht]
  \begin{center}
    \begin{tabular}{c}
      \includegraphics[height=8cm]{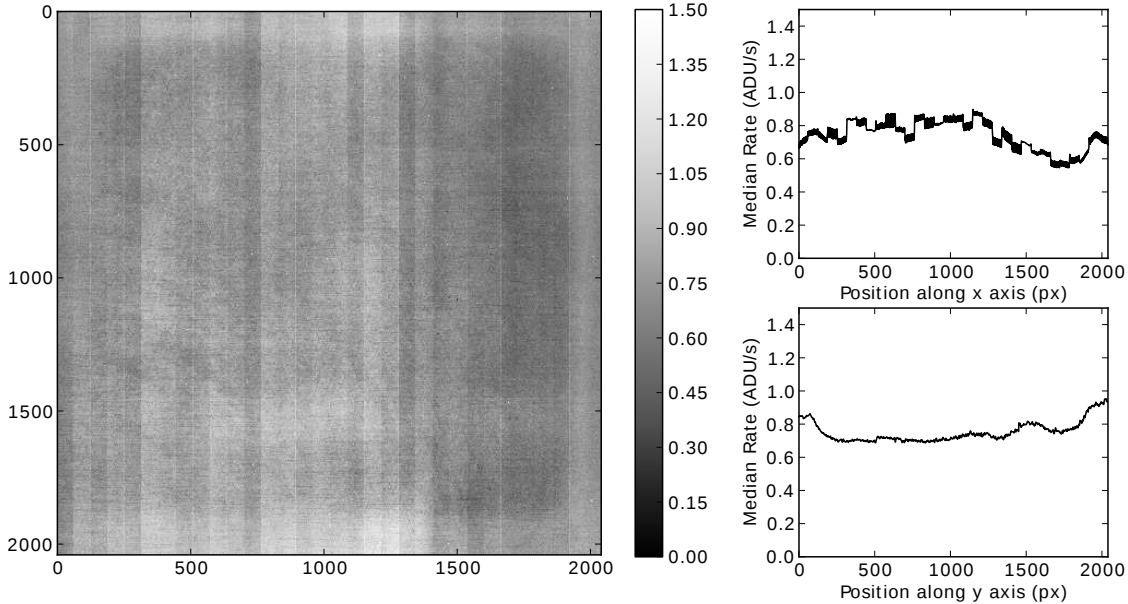}
    \end{tabular}
  \end{center}
  \caption 
  { \label{fig:dark_current}
  2D map of the rate of charge accumulation for each pixel in an UTR sequence. The panels on the right show the median rate along both collapsed axes. The effect of different DC offsets for each output channel is evident in the topmost right panel. Units are in ADU/s.
  }
\end{figure}

To quantitatively assess if the problem had been resolved, the rate of charge accumulation was plotted for each pixel in an UTR sequence following these modifications. A median rate of $<$1 ADU/s was calculated following the application of a four parameter fit discussed in \S \ref{sss:reset}. The full frame result is shown in Figure \ref{fig:dark_current}. Aside from the square imprint of the filter housing around the edge of the frame, no significant variation in gradient across the chip is observed.

\subsection{Cryogenic Considerations and Temperature Control}

In order to maintain temperature, SupIRCam had to be placed on the vacuum pump at intervals of between 2--4 weeks. This was thought to be due to the positioning of the getter chassis on the floor of the radiation shield where the temperature was found to plateau at around 160K. In order to make the pumping action of the getter more effective, it was required to attain a lower temperature as the residence time for a molecule adsorbed on the surface of a material, $\tau$, is approximately given by\cite{jousten1999thermal}:

\begin{equation}
 \tau = \frac{1}{\nu_{0}}\mathrm{exp}\left(\frac{E_{A}}{RT}\right)
\end{equation}
\\
\noindent where $\nu_{0}$ is the frequency of oscillation of a molecule on the surface (taken as $10^{-13}$s) and 
$E_{\textit{A}}$ is the activation energy of desorption. From Ref. \citenum{doi:10.1021/je100024r}, an $E_{\textit{A}} \approx 40\mathrm{kJ}/\mathrm{mol}$
is assumed (typical of physisorption binding energies), giving a residence time of the order of seconds at 160K. After repositioning the getter directly on the detector cold head, the temperature of the getter material now reaches about 90K. Consequently, the residence time of adsorbed molecules, and thus the vacuum hold time, should have increased significantly. In practice, however, other factors such as the initial conditions of the getter (and consequently adsorption capacity) make this time an unachievable upper limit of the actual vacuum hold time. The cryostat has been empirically found to have a hold time of about 8 months, although recent testing of a strategy to purge the getter material through baking and saturation with nitrogen is yielding promising results for future maintenance.

In order to servo the detector temperature, a LakeShore DT-471-CO silicon diode has been clamped to the detector mount (see Figure \ref{fig:invar_block}) alongside a single 25$\Omega$ resistor used as a heating element. Both the diode and resistor are connected to a Lakeshore model 331 controller, with the diode having both load and sensing lines. This configuration provides a stability of 4.2--5.5mK over a temperature range of 77--300K, with the resistor capable of outputting between 0.25W and 25W of heater power when connected to the LakeShore's primary control loop. To mitigate the possibility of exceeding a rate of temperature change greater than the recommended 1K/min, the heater power has been set to a maximum of 2.5W. A separate DT-471 built in to the SIDECAR is also connected to this controller to monitor the ASIC temperature, but is not connected to a control loop. As this generation of controller has only a serial interface, a custom transceiver with a built-in ethernet server is used to control it over the telescope network.

\subsection{Software and Instrument Control System Hardware}

The detector is controlled using software provided by Teledyne. Teledyne uses a third party USB driver (Bitwise Systems QuickUSB) and their own hardware
abstraction layer (HAL) to translate application commands into driver specific commands, hiding the USB driver details. In turn, a Teledyne supplied COM DLL object uses a .NET web service to communicate with the HAL. This COM DLL object allows third party software to interface with the HAL. Teledyne provide an example front-end interface to these services written in IDL. This interface has an inbuilt socket server, enabling control over TCP. For this interface to work, a special version of the assembly code (HxRG) must be uploaded to the SIDECAR. This code allows the user to change certain exposure options ``on-the-fly'' without having to reupload the assembly code. Such options include the number of detector outputs used during readout, preamp gain, windowing and the exposure mode (viz. UTR sampling, fowler sampling) together with their associated parameters.

As the USB driver is Windows only, IO:I uses two ICCs: one running Windows and one running a recent version of CentOS, a stable and predictable derivative of Red Hat Enterprise Linux. The windows machine is used to host the IDL socket server that communicates with the detector. The linux machine hosts a remotely accessible instrument agnostic interface which communicates with the socket server. This interface uses a command set that is homogeneous to the other instruments on the LT\cite{2004SPIE.5492..677M}, allowing for easy integration with the telescope's Robotic Control System (RCS). The linux machine also deals with other aspects of data management, such as renaming output files in accordance with LT conventions and adding the appropriate FITS headers relating to the telescope and observing conditions.

\subsubsection{A Note Regarding Virtualisation}

It has already been shown by RATIR\cite{2012SPIE.8453E..2SK} that this two machine setup lends itself well to virtualisation. For IO:I, therefore, these machines were initially hosted as guests under a linux host running VirtualBox, an open source virtualisation environment. A separate shared storage space within this environment was provided to both of the guests in order to store data taken with the instrument. However, two key problems were encountered during testing of this mode of operation: i) significant overhead on performing header keyword write/update routines provided by the CFITSIO library, and ii) seemingly random FIFO overflow errors during some exposure sequences.

The first of these issues arose from the need to append and update headers to the files outputted by the IDL software. Quantitatively speaking, it took approximately two seconds per file to process the required 72 header cards. Given that the process would return almost instantaneously on the native ext3 file system, this overhead is most likely a caching issue relating to the virtual filesystem (VFS) VirtualBox uses for shared folders. The problem was also found, albeit to a lesser extent, when using VMWare - a proprietary alternative to VirtualBox. Even for exposure sequences containing a small number of images, this delay becomes a significant contributor to the instrument usage overhead. 

The FIFO overflow errors are more problematic as they cause an abort of the exposure sequence when they occur. This problem is not exclusive to a virtualised setup, but is thought to be exacerbated by the additional overhead incurred by writing to the VFS. Other solutions were tested, including writing to a Network File System (NFS) share located elsewhere on the network, but this did not address the underlying cause and the problem persisted with a similar frequency of failure. 

To mitigate both of these issues and improve reliability, two separate physical machines are now used and the files are shared between them both through an NFS share hosted on the windows 
machine\footnote{http://www.hanewin.net/nfs-e.htm}. In this manner, the files are written locally to the drive's native filesystem, now hosted on a solid state disk with much greater Input/Output Operations Per Second (IOPS) than the previous spinning hard drive. The header write/update overhead has notably decreased by roughly an order of magnitude and, to this date (December 2015), no FIFO overflow issues have been seen to occur.

\subsection{Electrical}

\begin{figure}[ht]
  \begin{center}
    \begin{tabular}{c}
      \includegraphics[height=11cm]{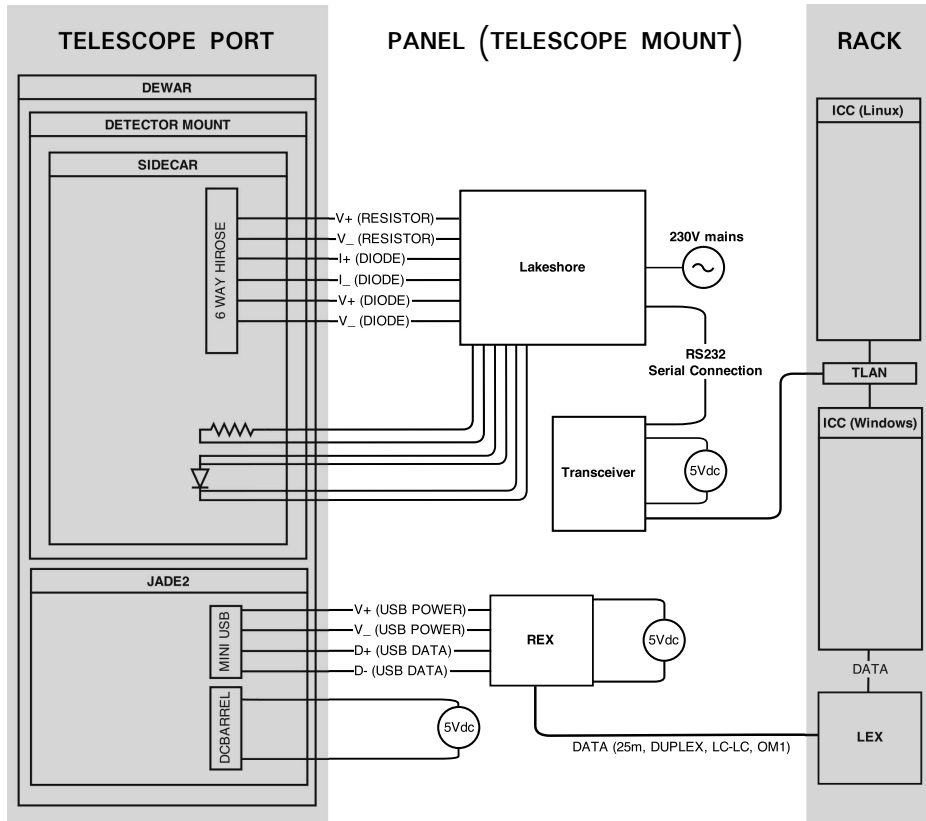}
    \end{tabular}
  \end{center}
  \caption 
  { \label{fig:wiring}
  Wiring schematic. REX and LEX refer to the remote and local fibre optic extender units respectively. TLAN refers to the telescope ethernet network.} 
\end{figure}

A broad overview of the electrical and optical connections both inside and outside the cryostat is shown in Figure \ref{fig:wiring}. The data lines for the JADE2 are coupled to a fibre optic transceiver, allowing the transfer of data from the telescope to an ICC which is rack-mounted in a room adjoined to the telescope enclosure.

\subsubsection{Noise and Grounding Considerations}

\begin{figure}[ht]
  \begin{center}
    \begin{tabular}{c}
      \includegraphics[height=9cm]{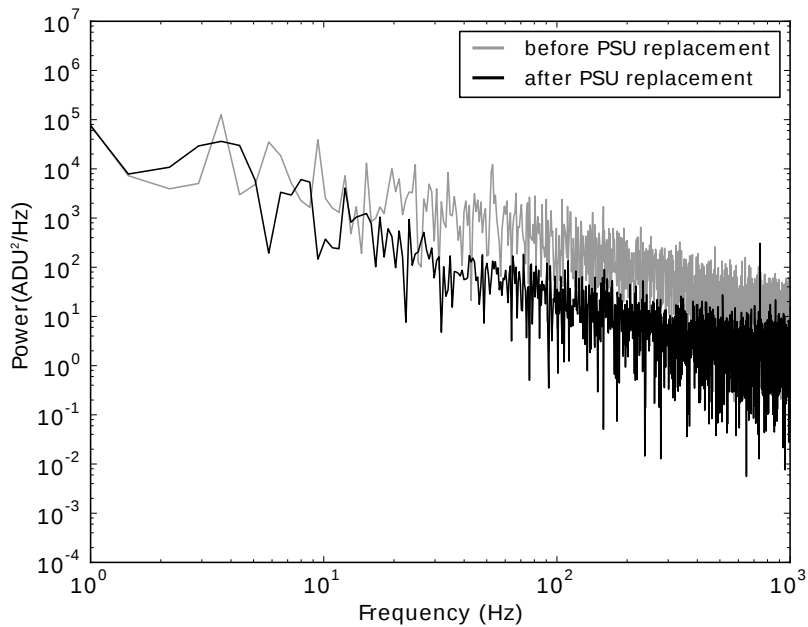}
    \end{tabular}
  \end{center}
  \caption 
  { \label{fig:noise_improvements_2}
  Noise power spectrum before and after both the faulty SIDECAR PSU was replaced and the switched-mode PSU powering the JADE2 USB was changed. The spike at 50Hz in the before PSU replacement plot is mains pickup, most likely the result of using a switched-mode PSU.} 
\end{figure}

To minimise voltage ripple, two linear power supply units (PSUs) provide power separately to both the SIDECAR and the USB section of the JADE2 circuit. A switched-mode PSU was initially used to power the JADE2 USB, but was found to introduce noticable mains pickup noise (see Figure \ref{fig:noise_improvements_2}). To avoid ground loops, all internal analog/digital grounds from both the SIDECAR and JADE2 boards (apart from the JADE2 USB) are tied to a single point ground provided by the SIDECAR PSU (see Ref. \citenum{2012SPIE.8453E..1OF} for specifics). A replacement PSU was sourced for the SIDECAR after the noise from the first unit was found to correlate with increasing temperature, producing vertically banded frames upon readout. 

\section{Detector Characterisation}

The following data was taken in UTR, Correlated Double Sampling (CDS) and Fowler observing modes. To minimise analog-to-digital converter (ADC) readout noise, the detector was operated at the slower pixel clocking speed of 100kHz using the Successive Approximation ADCs with a resolution of 16 bits. The output signals were measured differentially against a common reference, \textbf{InPCommon}, from the detector. The temperature of the array was held at 85K. 

\subsection{Dark Current}

During early testing of the array, a blank was used in place of a filter to enable true darks to be obtained. However, a combination of reset anomaly and an elevated charge accumulation rate made these frames unusable for the purpose of measuring dark current. Without empirical data, we must rely on the manufacturer's specification. Teledyne quote a median dark current for this array of less than 0.01e-/s at 120K. The magnitude of this figure is corroborated by RATIR who, with the same model array, were unable to measure any dark current for 1000s CDS exposures operating at a detector temperature of $\sim$60K\cite{2012SPIE.8453E..1OF}. Although we now cannot remove this component independently, the dark current, and indeed any other 2D imprint on the data (such as thermal glow from the telescope), is typically removed during the sky subtraction phase of data processing. However, if the pipeline-generated sky is a poor representation of the true sky -- as will be the case if the observation is of a sufficiently extended source -- then the observer will be required to generate their own offset sky and do this step of the reduction process themselves.

\subsubsection{Reset Anomaly}\label{sss:reset}
In all of the UTR sequences taken with the blank, an appreciable early time nonlinearity was found to affect all pixels in the array. This nonlinearity manifested as an additive component to the integrated signal, and took an approximately exponential functional form. The problem, known as ``reset anomaly'', has been observed previously in hybridised HgCdTe detectors and is thought to be due to charging effects in the readout integrated circult (ROIC)/detector\cite{2007PASP..119..768R}. Reset anomaly is normally only observed to affect a fraction of the pixels in the array, but the effect here presented in all pixels\cite{2014SPIE.9147E..37B}. Neglecting to take this effect into account leads to systematic overestimation when determining dark current.

A solution has previously been put forward to allow determination of a more accurate value for dark current when reset anomaly is present\cite{2004SPIE.5563...35B}. This solution involves determining the coefficients of a four parameter fit to the integrated signal as a function of time. These parameters describe both exponential and linear components, the latter of which can be used to calculate the dark current in an unbiased manner without having to arbitrarily estimate where the contribution of the exponential component becomes negligible. However, it is evident from the magnitude of the charge accumulation rates in Figure \ref{fig:dark_current} that the dark frames obtained for this analysis contain an additional diffuse component that uniformly illuminates all pixels in the array. This component is thought to be due to radiation originating from the blank and the surrounding filter housing. As the contribution from this component can be subtracted off during the data reduction process, this problem has not been investigated further and consequently no direct empirical measurement of dark current is possible.

\begin{figure}[ht]
  \begin{center}
    \begin{tabular}{c}
      \includegraphics[height=9cm]{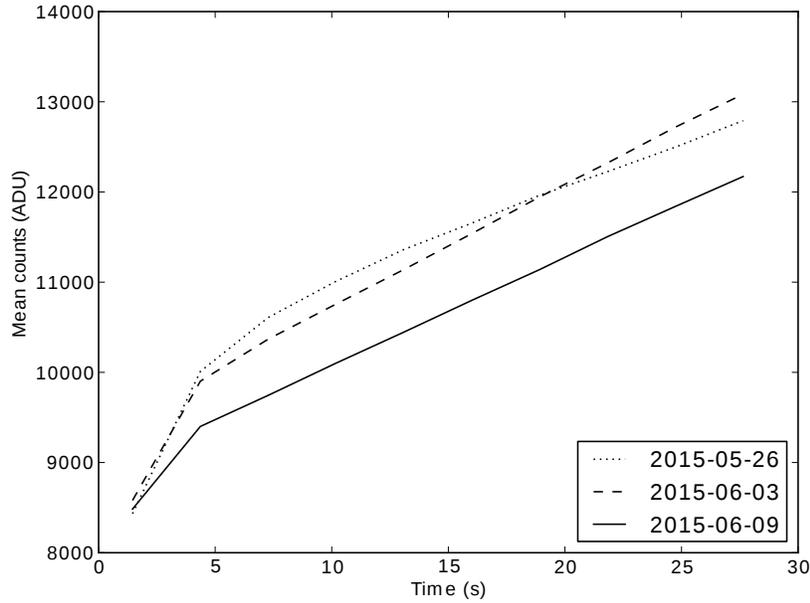}
    \end{tabular}
  \end{center}
  \caption 
  { \label{fig:reset_anomaly_dynamic}
  An illustration of how the reset anomaly is changing with time. Acquisition dates for each of the sequences are shown in the legend. In the most recently taken sequence (2015-06-09), the nonlinear jump in magnitude during the early part of the ramp is seen to be considerably smaller. The later part of sequences 2015-06-03 and 2015-06-09 are also more linear than 2015-05-26 after $\sim$5s. } 
\end{figure}

Additionally, the form and magnitude of the reset anomaly for this detector has been observed to change with time. Although the effect was not observable in any of the test data used in the following sections, it has presented again in UTR sequences taken at several dates following a recent thermal cycle. Plots of the mean count as a function of time are shown in Figure \ref{fig:reset_anomaly_dynamic}. The detector had settled at its operating temperature.

To ascertain if this effect was primarily additive (and thus fully correctable during data reduction), the photometric linearity tests discussed in \S\ref{ss:linearity} were performed on datasets both with and without reset anomaly. No measurable error on the former dataset could be detected.

\subsection{Gain and Read Noise}

\begin{figure}[ht]
  \begin{center}
    \begin{tabular}{c}
      \includegraphics[height=13cm]{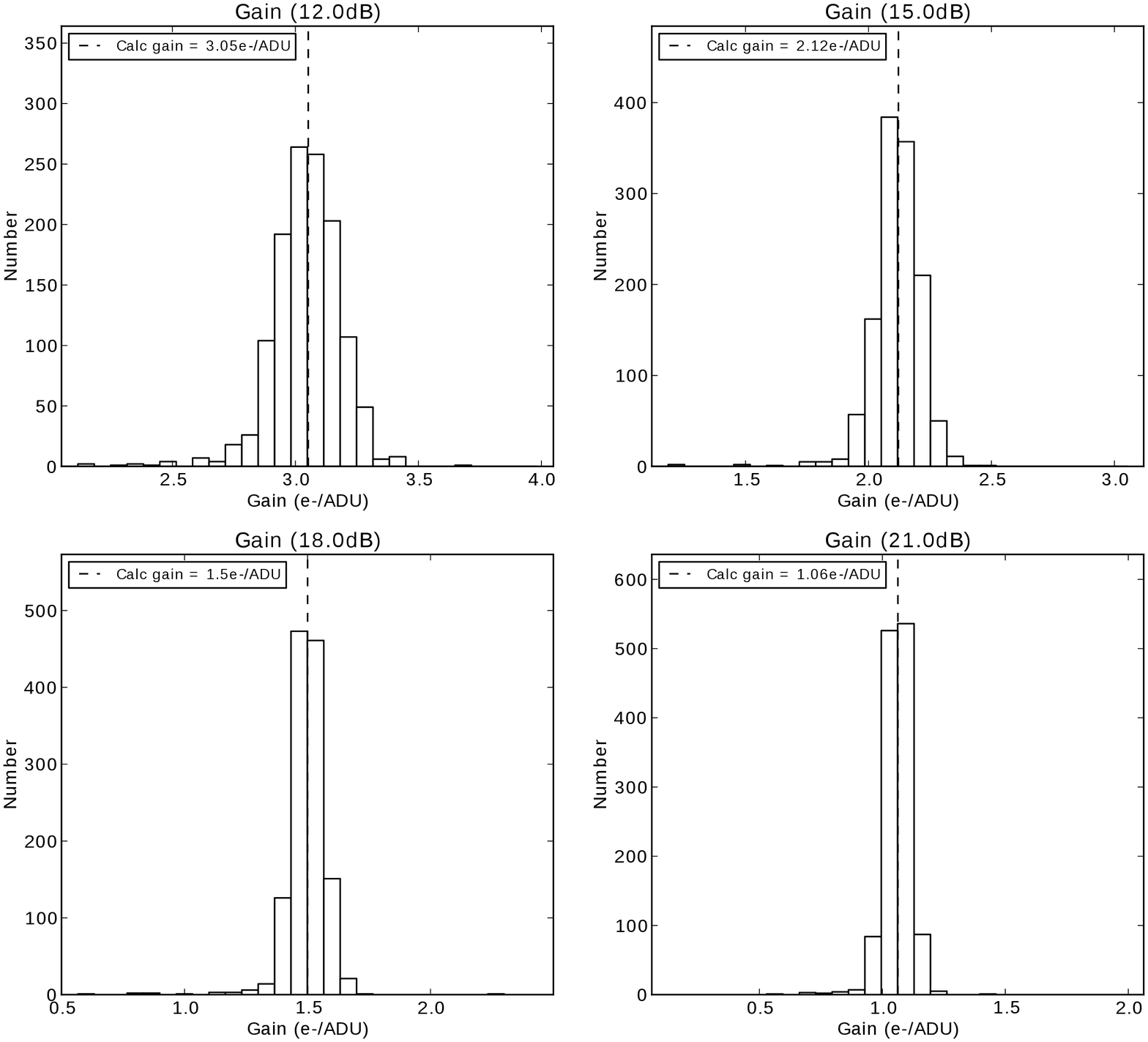}
    \end{tabular}
  \end{center}
  \caption 
  { \label{fig:gain}
  Conversion gain for the different preamp settings. Conversion gain decreases as preamp gain increases.
  } 
\end{figure}

\begin{figure}[ht]
  \begin{center}
    \begin{tabular}{c}
      \includegraphics[height=13cm]{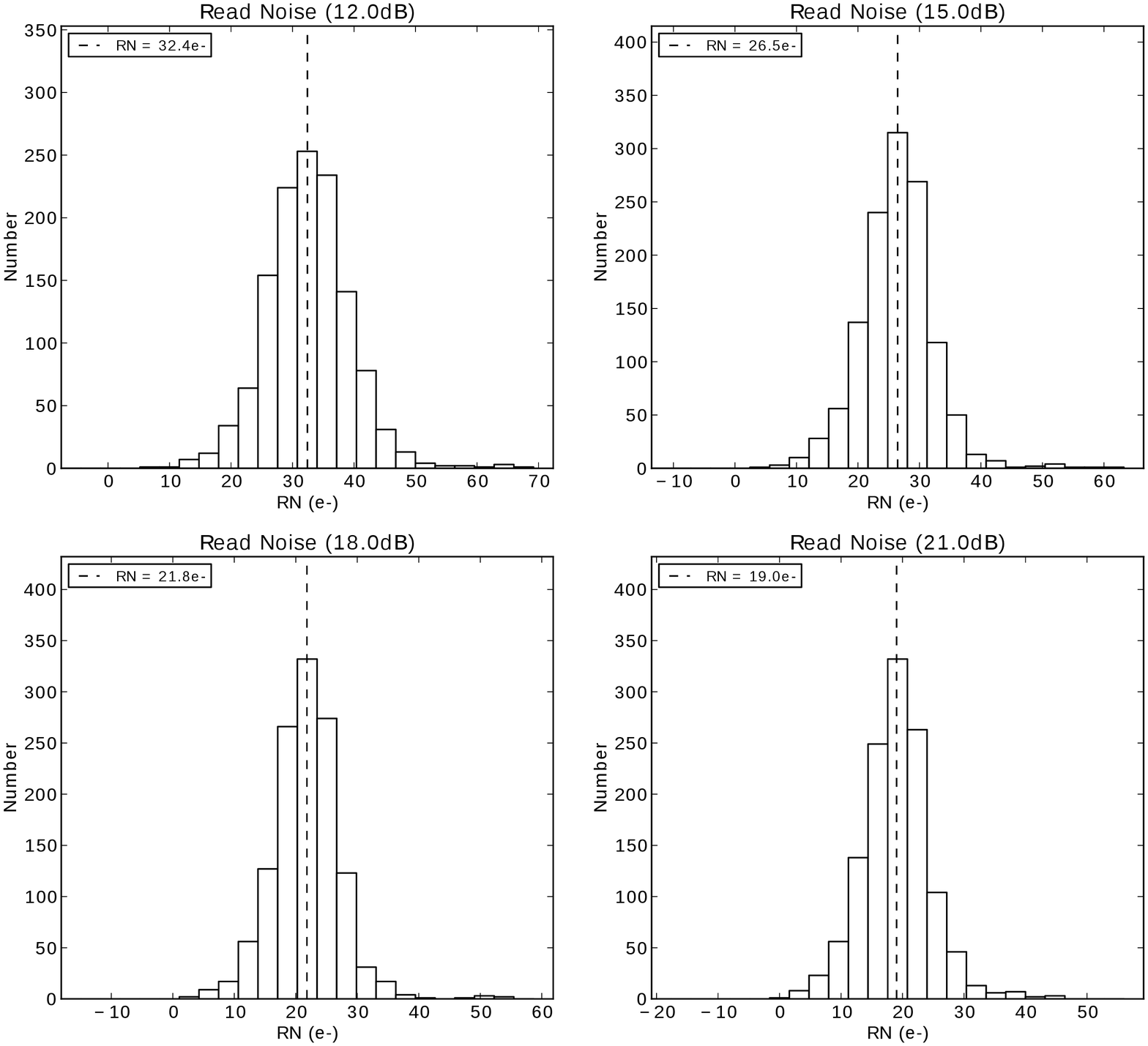}
    \end{tabular}
  \end{center}
  \caption 
  { \label{fig:read}
  CDS read noise for the different preamp settings. The read noise was converted into units of electrons by using the conversion gains determined previously. Read noise decreases as preamp gain increases.
  } 
\end{figure}

In the absence of true dark frames, the Photon Transfer Method \cite{janesick2007dn} was used to measure both read noise and gain for a range of preamp gains from 12dB to 21dB in 3dB steps. For each preamp gain, the reference voltage \textbf{Vrefmain} (SIDECAR register 0x602c) was adjusted so that the reference pixel signal from the detector was in range of the ADCs. The array was partitioned into 50x50 pixel windows and the gain/read noise calculated for each. A histogram of the results for each gain setting can be seen in Figure \ref{fig:gain} and \ref{fig:read}. 

For the four preamp gains of 12, 15, 18 and 21 dB, the four conversion gains attainable are 3.05, 2.12, 1.5 and 1.06 e-/ADU respectively.

The reference subtracted read noise calculated in Figure \ref{fig:read} is the CDS read noise, which is greater than read noise for a single frame by a factor of $\sqrt 2$. By observing in Fowler mode, the read noise can be further decreased by increasing the number of pairs in the observing sequence; this is illustrated in \S \ref{ss:num_fowler_reads}. For the four preamp gains of 12, 15, 18 and 21 dB, the four values of CDS read noise attainable are 32.4, 26.5, 21.8 and 19 e- respectively.

\subsection{Full Well Depth and Nonlinearity}

\begin{figure}[ht]
  \begin{center}
    \begin{tabular}{c}
      \includegraphics[height=11.4cm]{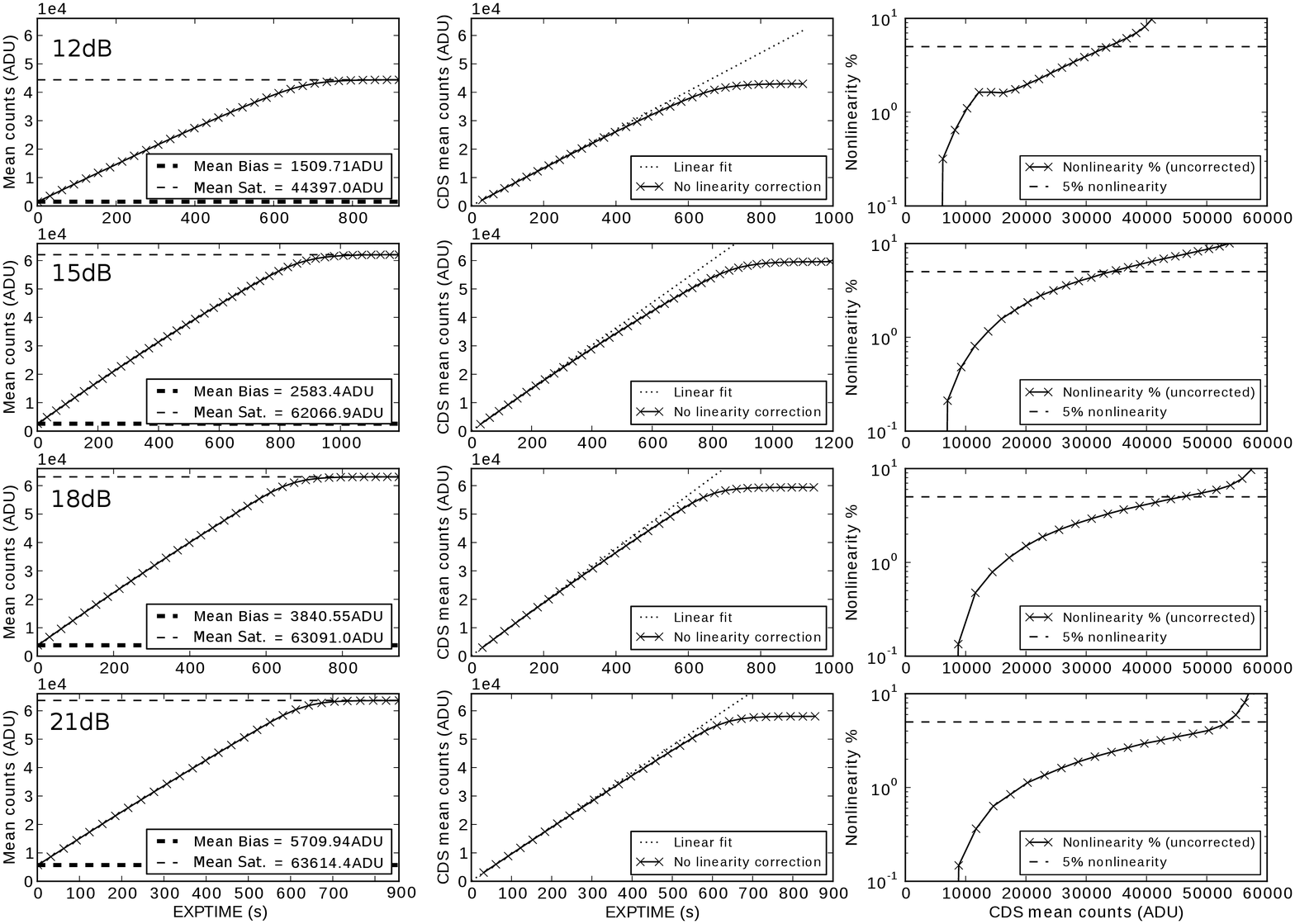}
    \end{tabular}
  \end{center}
  \caption 
  { \label{fig:fwd}
  Measurements of full well depth. The results for the different preamp settings are shown on different rows. The left panes show how the mean count increases for a region of the array that was free from cosmetic defects. The middle panes illustrate how the CDS mean count deviates from a linear fit to early integration times. The CDS mean count is taken as the difference between the first and last reads. The right panes show the percentage nonlinearity as a function of the mean count. 
  } 
\end{figure}

Full well depth (FWD) was measured by taking long UTR sequences that either saturated the detector or the ADCs. The results for each gain setting can be seen in Figure \ref{fig:fwd}. To match Teledyne's definition, FWD is hereinafter defined as the count at which nonlinearity $>$ 5\%.

For the four preamp gains of 12, 15, 18 and 21 dB, the four FWDs attainable without correction are 102, 72, 69 and 57 ke- respectively. Given that a higher preamp gain both reduces the read noise and full well depth, a compromise has had to be sought by distillation of the preceding analysis to the following statements:

\begin{itemize}\itemsep-5px
 \item[--] The 12dB setting underutilises the full range of the ADC and has the highest read noise.
 \item[--] The 21dB setting has a conversion gain of $\sim$unity, and is thus only capable of a FWD of $<2^{16}$e-. Given that a typical reduction sequence for the data is subject to both the removal of a DC bias level and CDS, this figure is actually an overestimate. Empirically this value has been found to be $\sim$57ke-.
 \item[--] Both 15dB and 18dB settings are viable. The 15dB setting has the advantage of only $\sim$3ke- larger FWD, but 18dB has $\sim$4.7e- less read noise. 
\end{itemize}

Favouring the reduction in read noise over increasing the FWD, the preamp gain has been set to 18dB.

\subsection{Nonlinearity Correction}\label{sss:nonlinearity_correction}

To account for nonlinearity, a series of calibration files were constructed using UTR sequences. This was done for only the 18dB setting. For each pixel, the integration time was plotted against the counts observed. A linear fit was then made to the early part of this sequence when the charge accumulation is assumed to be linear (no reset anomaly was present), and a third order polynomial fit made between the values extrapolated from the linear fit -- the ``actual'' counts -- and the observed counts. Per-pixel correction coefficients were then calculated. 

\begin{figure}[ht]
  \begin{center}
    \begin{tabular}{c}
      \includegraphics[height=9cm]{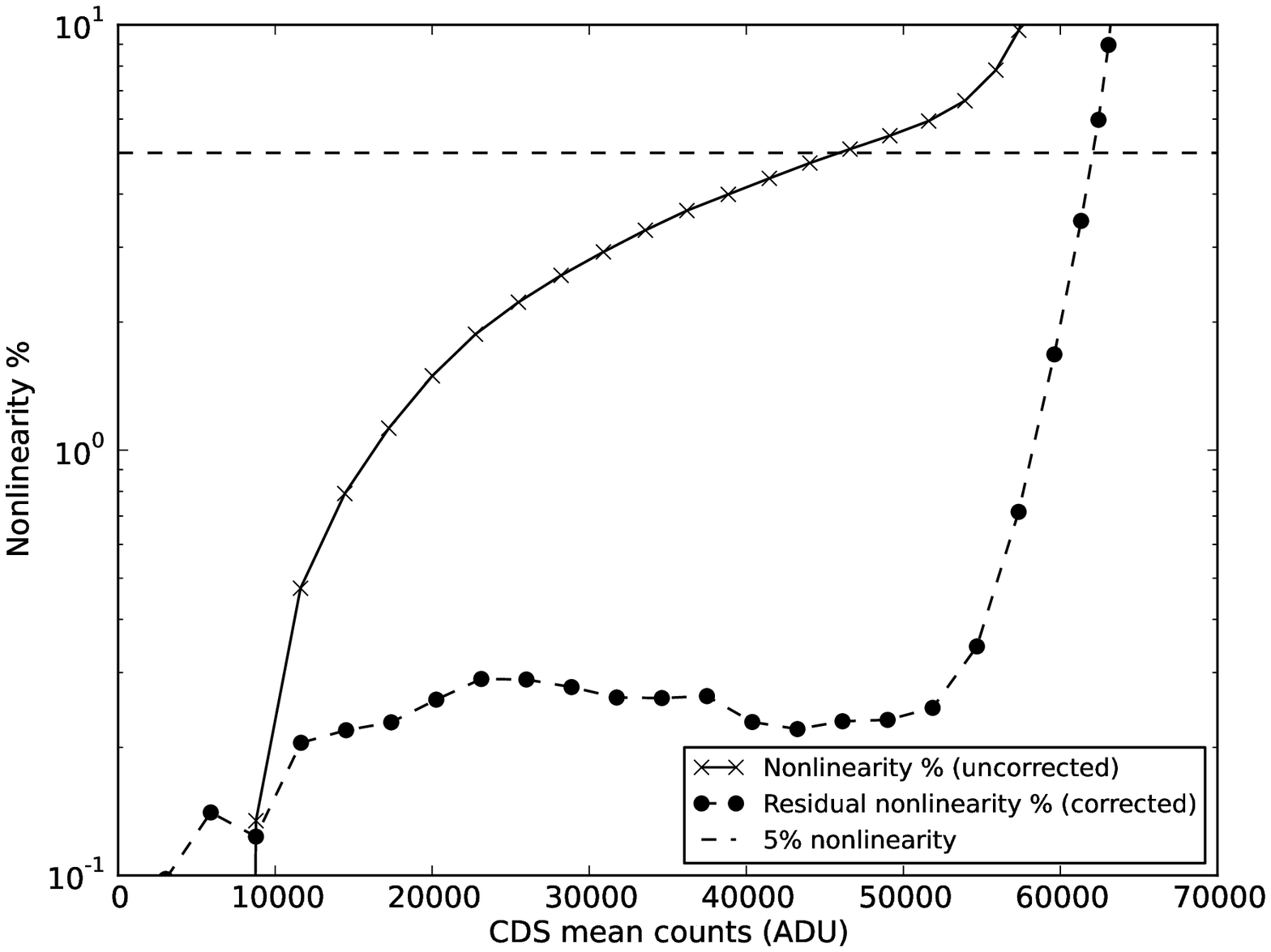}
    \end{tabular}
  \end{center}
  \caption 
  { \label{fig:linearity}
  Nonlinearity before and after correction for an UTR sequence at a preamp gain of 18dB.
  } 
\end{figure}

A correction derived from these coefficients has been applied to a separately obtained UTR sequence and the residual nonlinearity calculated. The result is shown in Figure \ref{fig:linearity}.

\subsection{Number of Fowler Pairs}\label{ss:num_fowler_reads}

\begin{figure}[ht]
  \begin{center}
    \begin{tabular}{c}
      \includegraphics[height=9cm]{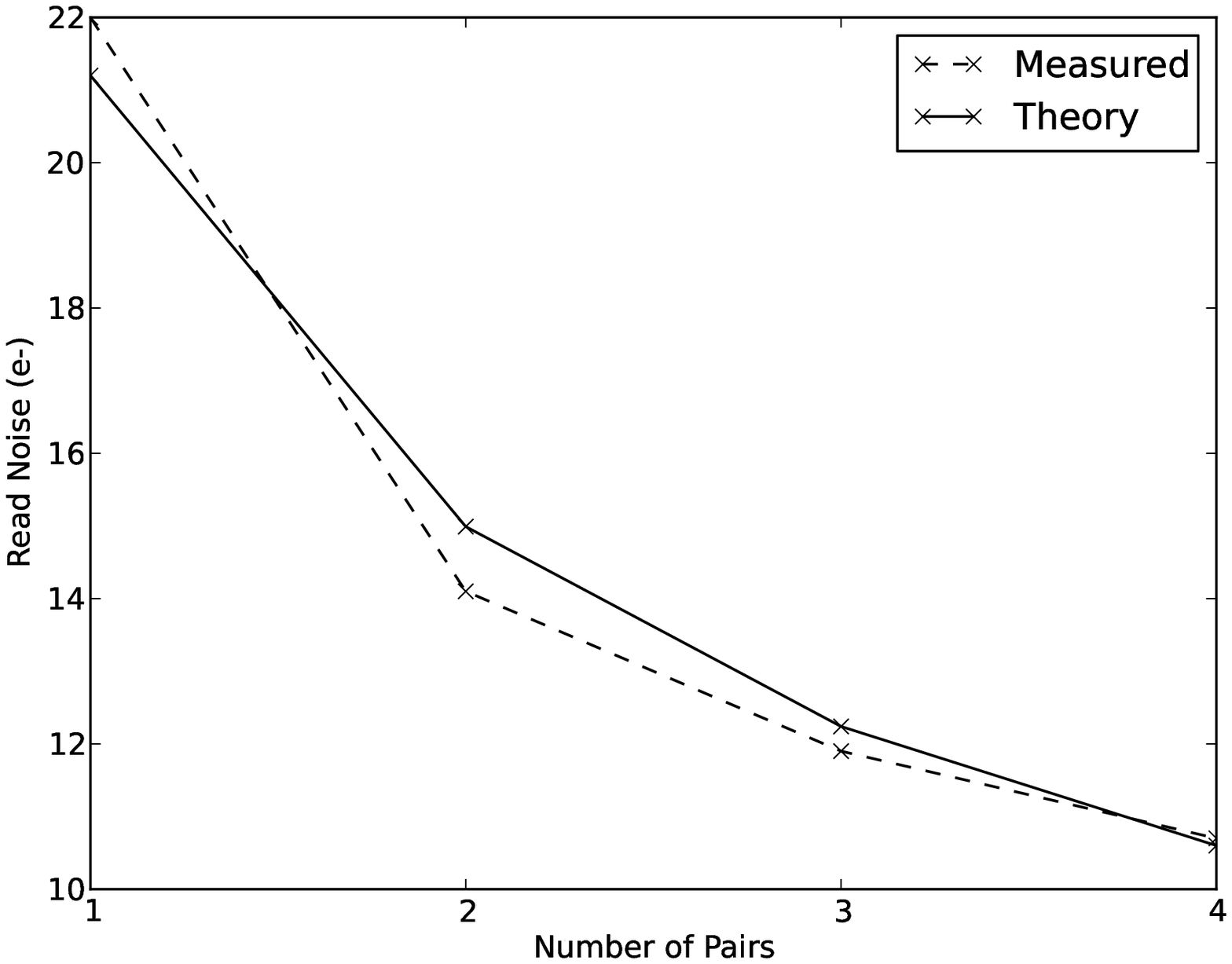}
    \end{tabular}
  \end{center}
  \caption 
  { \label{fig:fowler_pairs}
  Average read noise as a function of number of fowler pairs for a preamp gain of 18dB.
  } 
\end{figure}

For Fowler sampling, read noise is dependent on the number of pairs taken. The greater the number of pairs, the smaller the read noise that is attainable through averaging. This is shown in Figure \ref{fig:fowler_pairs}. This reduced read noise comes at the cost of increased observational overhead. For each additional pair, an extra read at the end of the sequence is required. Operating in 100kHz 32 output mode, a single extra read corresponds to an additional $\sim$1.3s. Increasing the number of pairs also has implications for data transfer from site and the resulting processing overhead. 

\subsection{Summary}

A summary of the preceding analysis is shown in Table \ref{tab:ioi_summary}.

\renewcommand{\arraystretch}{1.2}
\begin{table*}[ht]
 \centering
 \caption{A summary of IO:I's characterisation tests.}
  \begin{tabular}{l llll}
  \hline 
  \textbf{Preamp Gain} & \textbf{12} & \textbf{15} & \textbf{18} & \textbf{21} \\
  \hline 
  \textbf{Conversion Gain (e-/ADU)} & 3.05 & 2.12 & 1.50 & 1.06 \\
  \textbf{CDS Read Noise (e-)} & 32.4 & 26.5 & 21.8 & 19 \\
  \textbf{Uncorrected FWD 5\% (ke-)} & 102 & 72 & 69 & 57 \\
  \textbf{Corrected FWD 5\% (ke-)} & -- & -- & 93 & -- \\
  \textbf{Vrefmain (V)} & 1.71 & 1.26 & 1.16 & 1.09 \\
  \hline
  \end{tabular}
  \label{tab:ioi_summary}
\end{table*}

\section{Data Pipelining}

When used through the telescope's normal robotic interface, the instrument automatically sequences for the observer a dithered series of exposures to allow sky subtraction. 

\begin{figure}[ht]
  \begin{center}
    \begin{tabular}{c}
      \includegraphics[height=12.5cm]{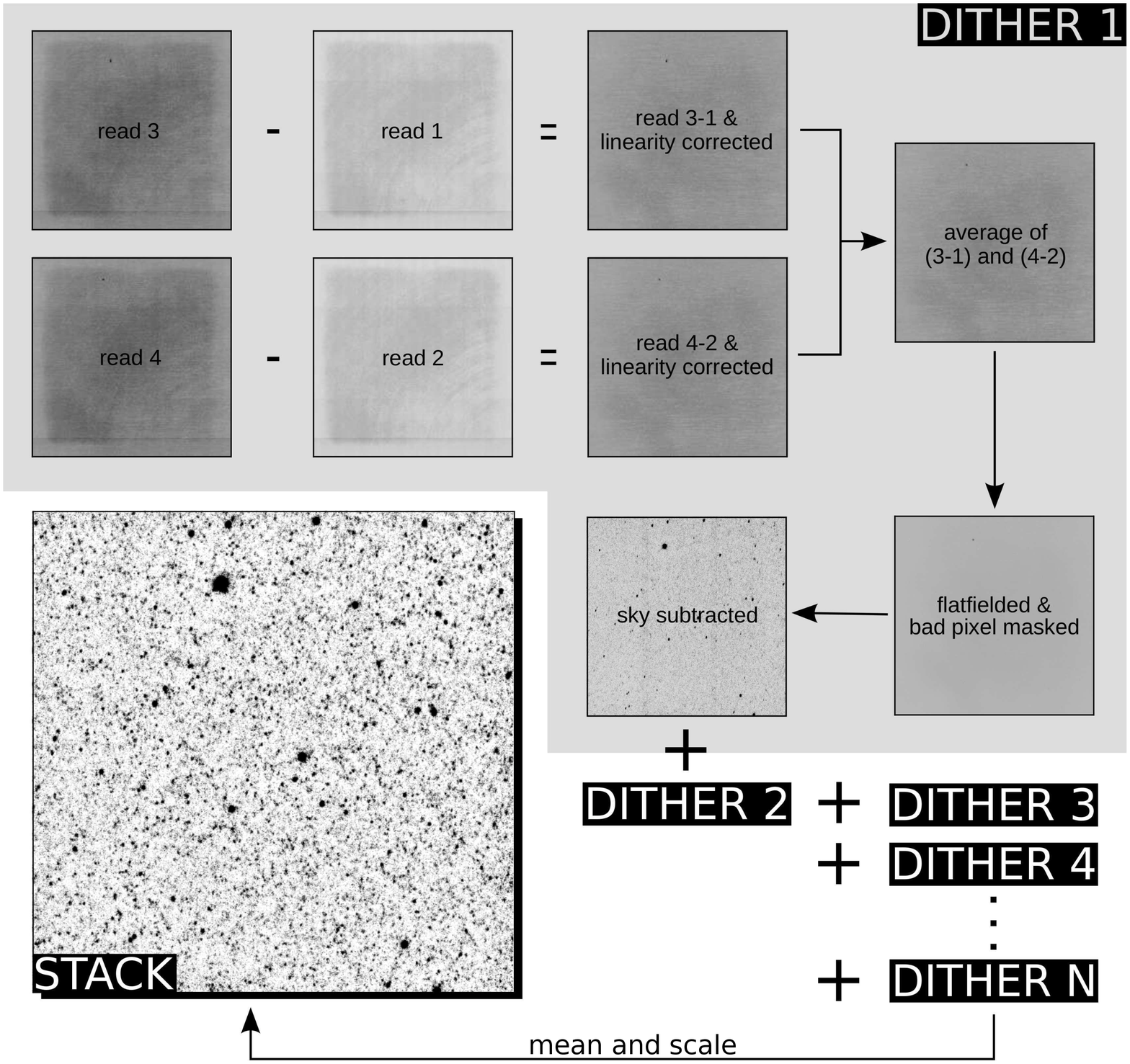}
    \end{tabular}
  \end{center}
  \caption 
  { \label{fig:datared}
  Example intermediate and end products from a typical reduction sequence using two pairs and \textbf{N} dithers. To make the stack, the dithers are averaged and scaled up by a factor of N.
  } 
\end{figure}

An autonomous pipeline has been developed to reduce data taken using this standardised observing sequence. This pipeline performs reference subtraction, frame combination, non-linearity correction, flatfielding, bad pixel masking, sky subtraction, registration and stacking. It is written in Python using PyFits and the third-party modules {\footnotesize\tt alipy}\footnote{http://obswww.unige.ch/$\sim$tewes/alipy/} and {\footnotesize\tt statsmodels}\footnote{https://github.com/statsmodels/statsmodels/}. An example reduction sequence is shown graphically in Figure \ref{fig:datared}. A brief description of the reduction recipe follows. 

\subsection{Reference Pixel Subtraction}

The active pixels of the H2RG array are bordered by a total of 8 columns and 8 rows of inactive pixels. Although these inactive pixels are not bonded to a photodiode, they are clocked during readout giving a means by which to reduce the common-mode noise present in active pixels. The simplest implementation of reference subtraction is to take an average of all of the inactive pixels and subtract this value from that of all active pixels. However, doing so does not take into account the cross-array, column-to-column and output-to-output drifts.

%

In an attempt to correct for these nuances, several different reference pixel subtraction schemes were tested. These schemes included, in varying combinations, single and per-output subtraction, fitting a linear slope to the upper and lower reference pixel regions (ramp) and distinguishing between odd-and-even columns (odd-and-even). For those schemes employing per-output subtraction, the array was split into 32 sections (i.e. the full 2048 pixel width of the array is split into sections of width 64 pixels, and each section is read through a different output amplifier). For those schemes employing ramp subtraction, the offset to be subtracted is a function of vertical position of the pixel in the array and was calculated on a per-row basis. A scheme to remove residual vertical banding by smoothing the columnar reference pixels is also currently being investigated. A comparison of the resulting read noise for each of the different methods is shown in Table \ref{tab:ref_sub_rn}.

\renewcommand{\arraystretch}{1.2}
\begin{table*}[ht]
 \centering
 \caption{Read noise for a variety of reference pixel subtraction methods. These values were calculated with a preamp gain of 18dB.}
  \begin{tabular}{ll}
  \hline 
  \textbf{Method} & \textbf{Read Noise (e-)} \\
  \hline
  None & 24.4 \\
  Single frame constant & 25.2 \\
  Per-output constant & 21.8 \\
  Per-output constant + odd-and-even & 21.8 \\
  Per-output ramp & 21.8 \\
  Per-output ramp + odd-and-even & 21.8 \\
  Per-output ramp + odd-and-even + vertical banding removal & 21.1 \\
  \hline
  \end{tabular}
  \label{tab:ref_sub_rn}
\end{table*}

In practice there is no measurable difference between subtracting a single per-output constant and fitting a ramp to the data. Distinguishing between odd and even columns also has no impact on the read noise. The pipeline currently uses per-output ramp + odd-even reference subtraction. More complex subtraction schemes do exist that make better use of these reference pixels, but these schemes have not been explored as they require alteration of the clocking pattern for readout of both reference and active pixels\cite{2011SPIE.8155E..0CR}.

\subsubsection{Correlated Noise}

\begin{figure}[ht]
  \begin{center}
    \begin{tabular}{c}
      \includegraphics[height=9cm]{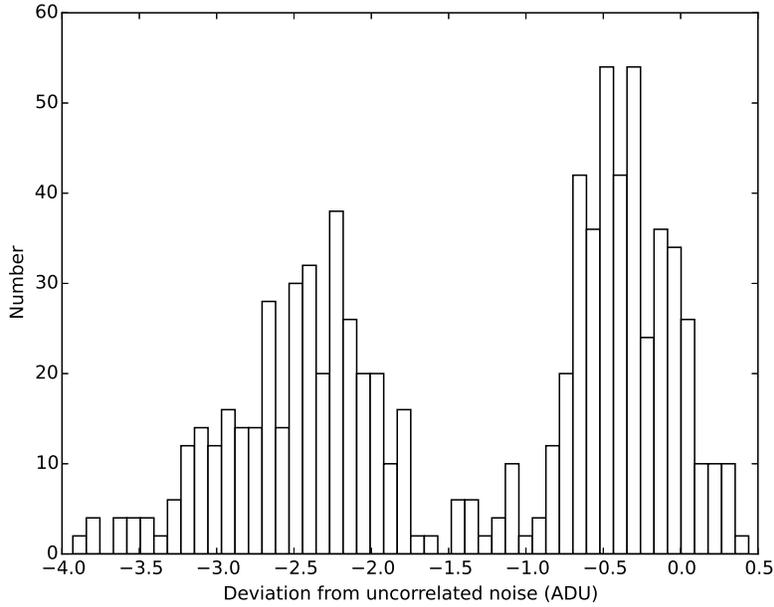}
    \end{tabular}
  \end{center}
  \caption 
  { \label{fig:correlated_noise}
  A histogram of the deviations between the expected theoretical noise of two differenced strips and the measured value. A negative value indicates that the measured noise was less than theoretically expected, indicating that correlated noise may be present. The bimodal form of the distribution indicates that only certain pairs of amplifiers are affected.
  } 
\end{figure}

Due to temporal drifts in the bias voltages supplied to the detector by the SIDECAR, there is the possibility of correlated noise between amplifiers. To quantify the extent this noise, we separated a short exposure image into 32 strips of data corresponding to one strip per amplifier. Each strip was then differenced with each other strip, and the standard deviation of this difference compared with the theoretical value expected for uncorrelated noise, given by taking the standard deviations of the two strips and adding them in quadrature. The distribution of expected theoretical noise relative to that which was measured is shown in Figure \ref{fig:correlated_noise}. There is evidence of a comparatively low level ($<$3 ADU) of correlated noise (cf. $\sim$24 ADU from Table \ref{tab:ref_sub_rn}). This issue may be further investigated and addressed in a later version of the pipeline.

\subsection{Frame Combination}

Both CDS and Fowler sampling are supported by the pipeline. As discussed in \S \ref{ss:num_fowler_reads}, Fowler sampling with two pairs is used for routine robotic operations.

\subsection{Nonlinearity Correction}

The calibration files discussed in \S \ref{sss:nonlinearity_correction} are used to correct for nonlinearity. Each order of the correction is stored in a separate extension of a single multi-extension file, allowing for easy application by array multiplication. 

A caveat exists to performing nonlinearity correction when the combination scheme is either CDS or Fowler sampling. In order for the nonlinearity correction factor to be calculated using an accurate reflection of the true charge stored in a pixel, it is required to first estimate the flux that is ``lost'' from the first readout due to taking the difference between pairs of frames. The timescale over which this flux is lost is equivalent to a multiple of the readout time. Calculating the rate of flux accumulation between the first and second frames of the first pair and factoring this into the linearity correction ensures that the correction is applied to the true value of pixel charge. If this lost flux is not taken into account, brighter targets (those with a higher rate of flux) will be systematically undercorrected.

\subsection{Flatfielding}

A variety of flatfielding techniques were considered but were deemed too problematic to obtain without a cooled blank. Without a blank, it is not easy to disentangle the QE of the detector and the dark current component. Doing so requires exposures of the same length but with different background brightnesses. Given this limitation, we have decided to use twilight difference images to construct the flatfield. There will inevitably be some discrepancy introduced from using twilight frames with a much higher colour temperature than typical of the sources being observed, but given that other schemes are not accessible to us presently, this is an unavoidable caveat. It should be noted that other flatfielding schemes, such as those using stacks of the night sky, may be ideal for flatfielding the night sky but would not necessarily match the colour of observed targets either.

To construct our master flatfield, a library of difference images was assembled using tracked and dithered twilight images of a blank field. Each image was processed through the normal pipeline as far as and including the Fowler combination. Difference images were then created using pairs of bright and faint twilight images of the same integration time. This removes all additive noise components such as overall dark current (expected to be negligible), dark current from a few hot pixels, reset anomaly and thermal background glow from the telescope itself -- typically of order 1kADU -- assuming that this glow is constant on short timescales (any change in this background would imprint on the flat). These difference images are level matched, combined with deviant pixel rejection and normalised to create the master flatfield.

\subsection{Bad Pixel Masking}

\begin{figure}[ht]
  \begin{center}
    \begin{tabular}{c}
      \includegraphics[height=11.5cm]{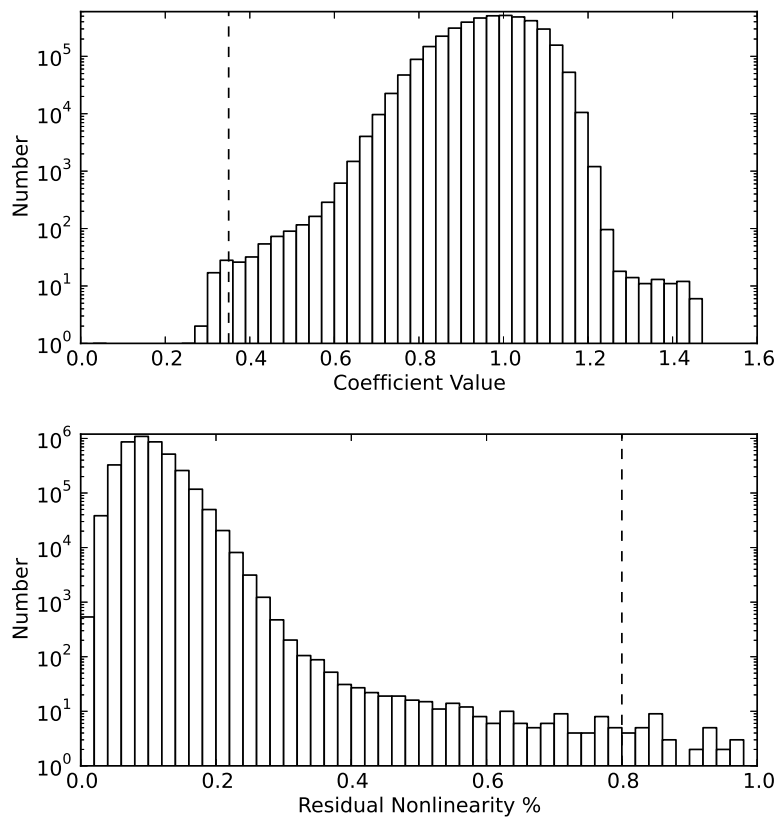}
    \end{tabular}
  \end{center}
  \caption 
  { \label{fig:bad_pixel_dist}
  Distribution of pixels as a function of flatfielding coefficient value (top) and residual nonlinearity (bottom). The dashed lines represent the lower (flatfielding) and upper (residual nonlinearity) limits set for a pixel to be flagged as ``bad''.
  } 
\end{figure}

By applying constraints to both flatfielding coefficients and nonlinearity residuals, pixel maps have been constructed to filter out bad pixels which either have a QE of $<$0.35 of the average and/or median residual nonlinearity of $>$0.8\%. The distribution of these two parameters is shown in Figure \ref{fig:bad_pixel_dist}.

\subsection{Sky Subtraction}\label{ss:sky_subtraction}

Sky subtraction is currently achieved by construction of a median stack using all dither positions \textit{except the one for which the stack is being generated} (``peers-only''). At the cost of increased sky noise, peers-only combination reduces the likelihood of systematically overestimating the central bright regions of each stellar PSF when using lesser robust estimators such as the median. An iterative sigma clip is then applied to determine and match the background levels of both the dither position being considered and the stack. The background-level-matched stack is then subtracted off.

The sky subtraction stage of the pipeline is something of a misnomer, insomuch that this stage also removes the contribution from both dark current and thermal glow.

\subsubsection{Unbiased Estimation of the Sky Value}

Other more robust M-Estimators were considered and tested, including Tukey's Biweight and Huber's T. Both of these algorithms use an iterative maximum likelihood solver that strives to select the ``average'' value that is most consistent with the sample values being drawn from a symmetrical, normal distribution. In the context of this particular application, finding the average value is equivalent to estimating the value of the sky at a given pixel. Like the mean, all data are considered in the final average but instead of considering each data point with equal weight, the data are multiplied by a weighting function that down-ranks points that lie further from the determined centre of the distribution. 

\begin{figure}[ht]
  \begin{center}
    \begin{tabular}{c}
      \includegraphics[height=8cm]{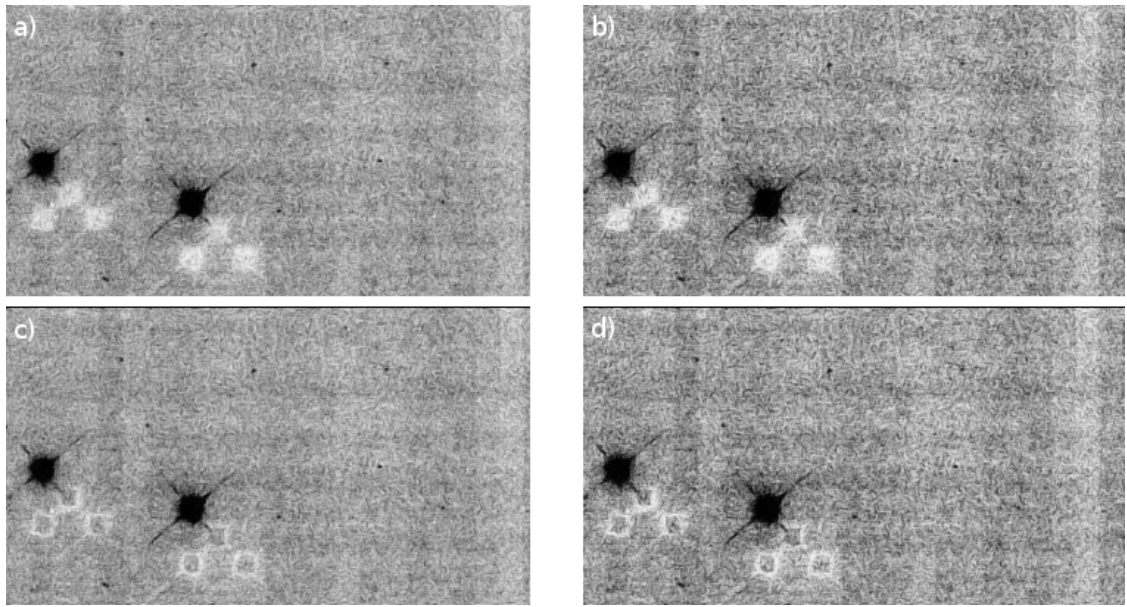}
    \end{tabular}
  \end{center}
  \caption 
  { \label{fig:robust}
  Illustration of residual artifacts post sky subtraction formed when using both a simple median and Tukey's Biweight to generate the stacked sky. The example is from a 4 dither pattern. Panels \textbf{a)} and \textbf{b)} are stacks generated using a median. Panels \textbf{c)} and \textbf{d)} are stacks generated using Tukey's Biweight. Panel \textbf{a)} and \textbf{c)} were generated using all frames. Panels \textbf{b)} and \textbf{d)} were generated using the peers-only scheme described in \S \ref{ss:sky_subtraction}. The sky noise in panels \textbf{a)}, \textbf{b)}, \textbf{c)} and \textbf{d)} is 37.7, 51.1, 36.6 and 49.4 ADU respectively.
  } 
\end{figure}

Initial testing of Tukey's Biweight looked to be the most promising alternative, proving much more robust to outliers than using a median or Huber's T. However, when applied to real data it was found that while the algorithm provides a better esimator of the sky background at the centre of the star PSFs the outer haloes of the stars were oversubtracted resulting in ringed artefacts (see Figure \ref{fig:robust}).

With careful consideration of aperture size, Tukey's Biweight yields the best absolute photometry. However, the artefacts created were thought to unnecessarily introduce confusion for users who were unfamiliar with this algorithm. This is especially true when one considers that the difference between the various subtraction schemes is significantly less noticable when an offset pattern utilising a greater number of dithers is selected. Consequently, the use of Tukey's Biweight was rejected for common use in favour of a simple median. 

\subsection{Image Registration and Stacking}

The pipeline uses the {\footnotesize\tt alipy} module to find the per-frame tranformation matrix with components of translation, rotation and scale. The first dither position is selected as a reference by which all the other dither positions are registered to.

\section{Photometric Tests of the Pipeline and Instrument Performance}\label{sec:phot}

The best test of the reduction pipeline and real world instrument accuracy is the consistency and repeatability of photometric measurements on sky. For this analysis, four fields were selected and observed with various integration times ranging from 6 to 60 seconds. Though the selected fields were centred on stars from the 'UKIRT Fundamental and Extended Lists'\cite{2001MNRAS.325..563H}, all 7 $<$ H $<$ 16 stars in the 2MASS Point Source Catalog\cite{2006AJ....131.1163S} were included in the analysis unless otherwise stated.

\subsection{Linearity and Sky Brightness}\label{ss:linearity}

\begin{figure}[ht]
  \begin{center}
    \begin{tabular}{c}
      \includegraphics[height=11.0cm]{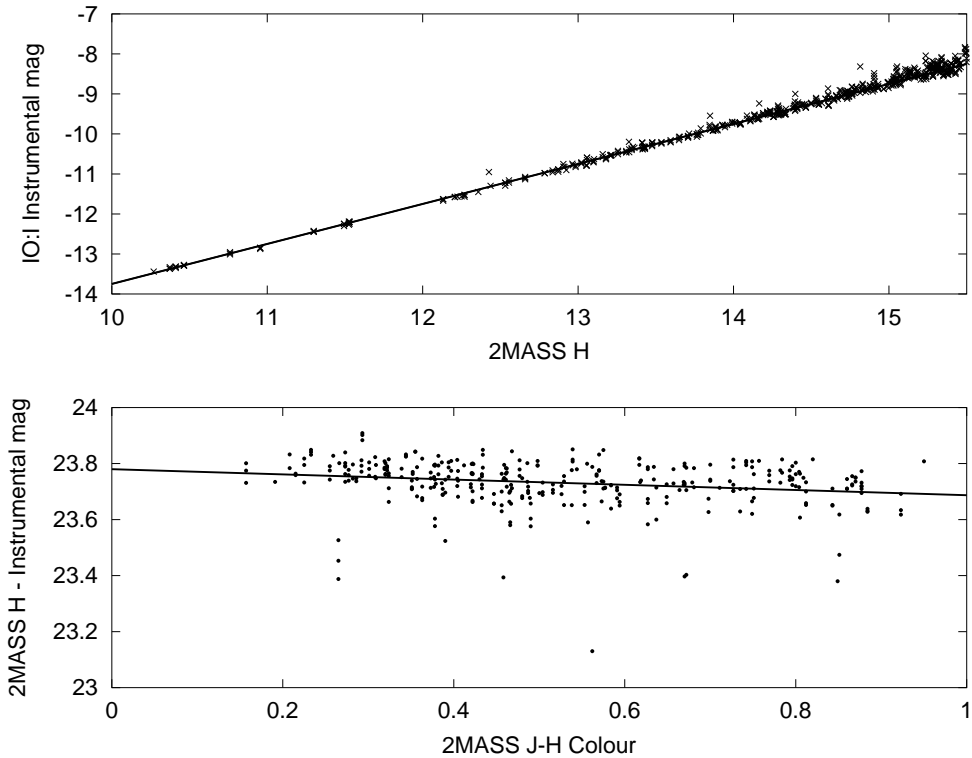}
    \end{tabular}
  \end{center}
  \caption 
  { \label{fig:phot_performance_lin}
  Top: Instrumental magnitude of sources calculated using frames taken with IO:I compared with corresponding values taken from the 2MASS catalogue. The overplotted straight line is not a fit, but simply a gradient of unity which would result from using a perfectly linear detector. Free fits to the data from different fields yielded gradients in the range 0.996 to 1.004. Bottom: Colour transformation with respect to 2MASS colour.
  } 
\end{figure}

To assess photometric linearity, observations of these fields were run through the IO:I pipeline and sources extracted using the Source Extractor (SExtractor\cite{1996AAS..117..393B}) software with several different circular apertures between six and twenty pixels diameter. Counts were converted to photo-electrons per second using a gain of 1.5e/ADU and instrumental magnitudes calculated. Figure \ref{fig:phot_performance_lin} shows a collation of these instrumental magnitudes on three different nights for the field surrounding standard star FS150, plotted against 2MASS PSC H-band magnitudes. The gradient of the fit to the data is consistent with unity, signifying the detector has responded linearly in these observations.

Since the detector red cut-off lies within the H-band it is to be expected that there will be a colour transformation between IOI:I H-band and other instruments. In practise, the transformation is not large for typical stellar sources though it could become significant for any particular source with strong spectral features between 1700 and 1800nm. Figure \ref{fig:phot_performance_lin} also shows the colour transformation with respect to 2MASS colour has a slope of -0.093 $\times$ (J-H) for all stars in the FS150 field.

The magnitude of the sky background was also calculated in the same analysis. We report a value of 13.34$\pm$0.08 mags/arcsec$^2$ -- about a half a magnitude brighter than the values quoted by both TNG\footnote{http://www.tng.iac.es/instruments/nics/imaging.html} and INT/JKT\footnote{http://www.ing.iac.es/Astronomy/observing/conditions/skybr/skybr.html}. This is not surprising given that no NIR optimisation has been made to either the telescope's infrastructure or environment. Indeed, there is a difference of $\sim$0.3 mags/arcsec$^2$ between the centre of the field and the outside, presumably the result of residual thermal glow.

\subsection{Consistency of Invidual Frames and Stacked Products}
\begin{figure}[ht]
  \begin{center}
    \begin{tabular}{c}
      \includegraphics[height=9cm]{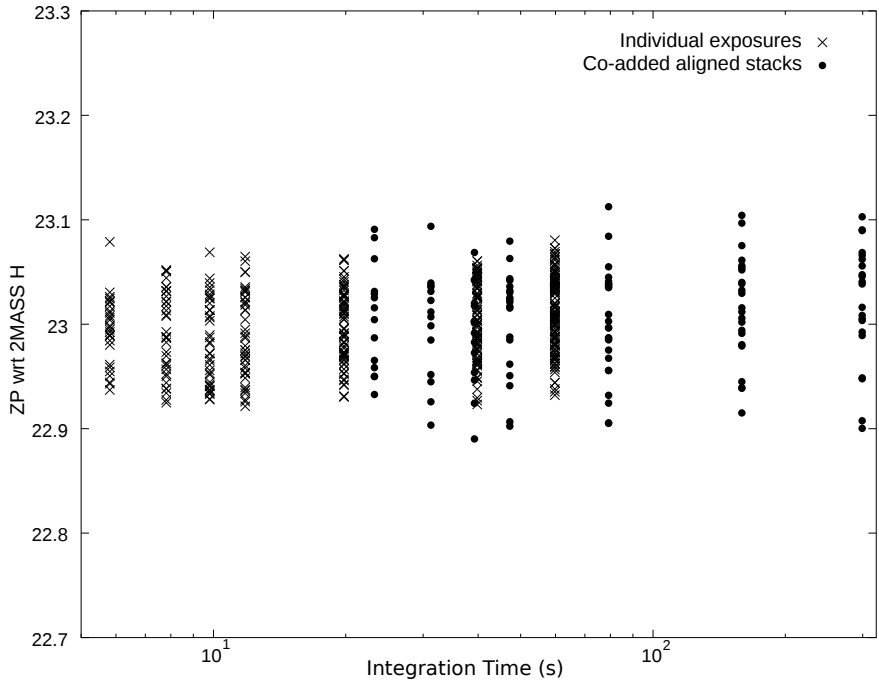}
    \end{tabular}
  \end{center}
  \caption 
  { \label{fig:phot_performance_exp_duration}
  Zeropoints calculated from both individual exposures and the stack as a function of exposure time. The zeropoint is defined to be 1ADU/sec referenced to the 2MASS PSC natural system and with no colour transformations.
  } 
\end{figure}

To check that the aligned, co-added data are stacked correctly in the pipeline, photometry was extracted independently for the same stars from each individual frame and from the final stacked data product. The zeropoints derived for all stars in two image fields are plotted against exposure time in Figure \ref{fig:phot_performance_exp_duration} for integrations ranging from 6 to 60 seconds and their associated five-frame-stacks. Photometry was taken from the sky-subtracted pipeline images with faint and saturated stars excluded.

\subsection{Sky Subtraction}
\begin{figure}[ht]
  \begin{center}
    \begin{tabular}{c}
      \includegraphics[height=9cm]{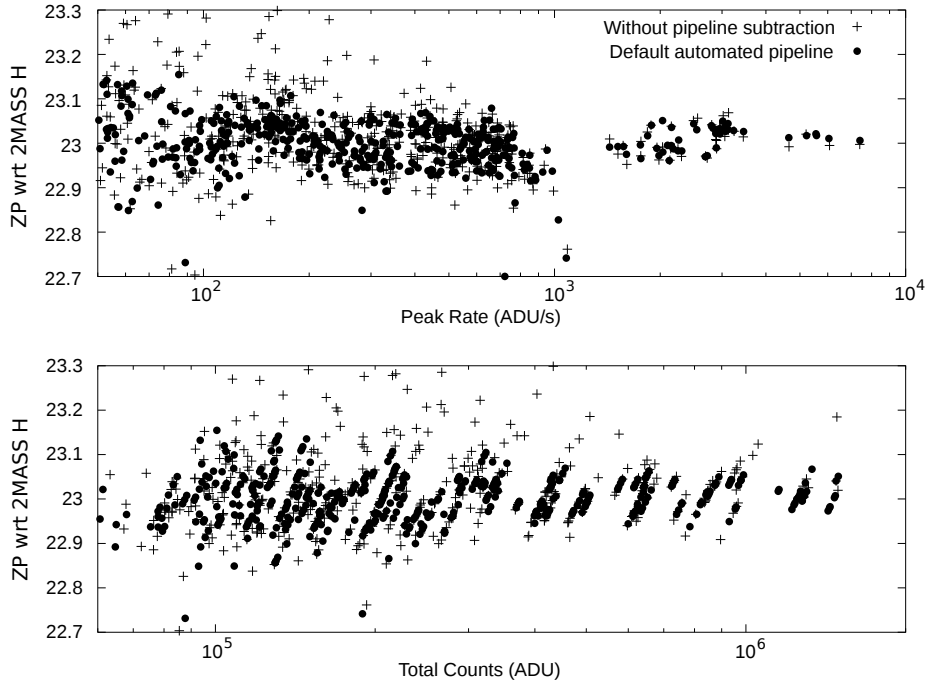}
    \end{tabular}
  \end{center}
  \caption 
  { \label{fig:phot_performance_skysub}
  Peak count rates and total counts as a function of zeropoint calculated with both default pipeline sky subtraction and without (instead using SExtractor sky subtraction). The striations in the lower panel are the result of using several exposures of the same field. Each group of points represents many observations of the same star. These groups form tilted lines because the 2MASS comparison reference which the observed counts are compared against is a single catalogue value whereas there is numerical scatter between our multiple observations.
  } 
\end{figure}

To test the pipeline sky subtraction and ensure that star residuals were not imprinted on the derived sky image, photometry was extracted from both the sky-subtracted image (FITS extension IM-SS) and the non-sky-subtracted data product (FITS extension IM-NONSS). Photometry was performed in SExtractor using both simple circular apertures and the "AUTO Kron-like elliptical aperture" mode. The two extraction modes gave consistent results. For the sky-subtracted image no sky subtraction was performed in SExtractor. For the non-sky-subtracted image, sky subtraction was performed using the package's default filtered sky background algorithms. Figure \ref{fig:phot_performance_skysub} compares the ensemble results for six different integration times on two different star fields. The derived ZP with respect to 2MASS is plotted both against total counts in the extraction aperture and against peak brightness at the centre of the star. Results are seen to be consistent for the two datasets and the scatter significantly reduced by using the pipeline's automated sky subtraction. This is expected since the pipeline makes use of multiple stacked images (see \S \ref{ss:sky_subtraction}) to derive a clean sky image with all the stars and other noise sources such as hot pixels removed.

\subsection{Persistence}
Image persistence from bright sources into subsequent images is detectable. A dithered sequence of images centred on a heavily saturated star with estimated peak counts of $1.3 \times 10^6$ADU shows a persistent image of peak 350ADU one minute after the initial exposure, fading to 170ADU two minutes after the exposure but undetectable after four minutes. Persistence is therefore not normally expected to be a significant problem and even one minute immediately after the saturation event the residual is less than 0.03 per cent, likely smaller than many other sources of photometric error. It is noted that using this analysis of persistence doesn't entirely probe the mechanisms whereby charge is trapped, which has been shown to be dependent more strongly on exposure cadence rather than intensity\cite{doi:10.1117/12.789372}. However these tests were performed with 60 second integrations, the longest recommended given the sky brightness and thermal background, thus representing an upper limit of persistence expected in normal science operations.

\section{Conclusions}

In this paper, the mechanical, electronic and cryogenic aspects of IO:I's development have been presented. This was followed by a discussion of results derived from characterisation tests, including measurements of read noise, conversion gain, full well depth and linearity. These measurements predicated the selection of 18dB as the most appropriate preamp gain, with corresponding parameters given in the 18dB column of Table \ref{tab:ioi_summary}. An overview of the data pipelining process was given and the results of photometric tests conducted on on-sky, pipelined processed data were presented. These tests have proved that the pipeline produces consistent and repeatable data products in terms of photometric linearity, both in sky-subtracted and non-sky-subtracted frames. 

Observers interested in soliciting a proposal to use IO:I are encouraged to apply through the website at http://telescope.astro.ljmu.ac.uk/PropInst/Call/. To assess programme suitability, an exposure time calculator is available at http://telescope.astro.ljmu.ac.uk/TelInst/calc/.

\acknowledgments 
The Liverpool Telescope is operated on the island of La Palma by Liverpool John Moores University in the Spanish Observatorio del Roque de los Muchachos of the 
Instituto de Astrofisica de Canarias with financial support from the UK Science and Technology Facilities Council. PyFITS is a product of the Space Telescope Science Institute, which is operated by AURA for NASA. The authors would like to thank both the anonymous referees for their insightful comments. RMB would like to thank Ori Fox, Bernard Rauscher, Craig Cabelli, Jing Chen and Dan Walters for their helpful advice and discussions. 

\bibliography{report}
\bibliographystyle{spiejour}

\end{spacing}
\end{document}